\documentclass[]{aastex631}

\usepackage{color}

\newcommand{\g}{\textbf}

\newcommand{\isois}{IS$\odot$IS }

\newcommand{\BB}{\bm{B} }

\newcommand{\Rsun}{$R_\odot$}

\shorttitle{Magnetic switchback occurrence}
\shortauthors{Pecora et al.}
\graphicspath{{./}}
\begin{document}

\title{Magnetic Switchback Occurrence Rates in the Inner Heliosphere: Parker Solar Probe and 1~au}

\author[0000-0003-4168-590X]{Francesco Pecora}
\affiliation{Department of Physics and Astronomy, University of Delaware \\ Newark, DE 19716, USA}
\email{fpecora@udel.edu}

\author{William H. Matthaeus}
\affiliation{Department of Physics and Astronomy, University of Delaware \\ Newark, DE 19716, USA}

\author[0000-0001-7004-789X]{Leonardo Primavera}
\affiliation{Dipartimento di Fisica, Università della Calabria \\ 87036 Arcavacata di Rende, Italia}

\author[0000-0001-5680-4487]{Antonella Greco}
\affiliation{Dipartimento di Fisica, Università della Calabria \\ 87036 Arcavacata di Rende, Italia}

\author[0000-0002-7174-6948]{Rohit Chhiber}
\affiliation{Department of Physics and Astronomy, University of Delaware \\ Newark, DE 19716, USA}
\affiliation{Heliophysics Science Division, NASA Goddard Space Flight Center \\ Greenbelt, MD 20771, USA}

\author[0000-0002-6962-0959]{Riddhi Bandyopadhyay}
\affiliation{Department of Astrophysical Sciences, Princeton University \\ Princeton, NJ 08544, USA}

\author[0000-0001-8184-2151]{Sergio Servidio}
\affiliation{Dipartimento di Fisica, Università della Calabria \\ 87036 Arcavacata di Rende, Italia}



\begin{abstract}

The subject of switchbacks, defined either as large angular deflections or polarity reversals of the magnetic field, has generated substantial interest in the space physics community since the launch of Parker Solar Probe (PSP) in 2018. Previous studies have characterized switchbacks in several different ways, and have been restricted to data available from the first few orbits. Here, we analyze the frequency of occurrence of switchbacks per unit distance  for the first full eight  orbits of PSP. In this work, are considered switchback only the events that reverse the sign of magnetic field relative to a regional average.  A significant finding is that the rate of occurrence falls off sharply approaching the sun near 0.2~au (40~\Rsun), and rises gently from 0.2~au outward. The analysis is varied for different magnetic field cadences and for different local averages of the ambient field, confirming the robustness of the results. We discuss implications for the mechanisms of switchback generation. A publicly available database has been created with the identified reversals.
\end{abstract}

%Use showkeys class option if keyword

%\tableofcontents

\section{Introduction}
\label{sec:intro}

The orbit of Parker Solar Probe (PSP) has taken the mission into previously unexplored regions
of the heliosphere in which both new phenomena, related to the origin of the solar wind, and known ones, from a rather different perspective, can be observed \citep{fox2016solar}.
The phenomenon of  magnetic field ``switchbacks'' (SBs), defined either as large deflections, or polarity reversals relative to ambient conditions, has been previously observed at larger distances \citep{Borovsky16,HorburyEA18}, but it is more evident in the PSP environment close to the sun.

A number of studies prior to PSP had already noted the existence of switchbacks in the interplanetary magnetic field \citep{McCrackenNess66, NeugebauerGoldstein13, Borovsky16, HorburyEA18}. The origin of these switchbacks or ``folds'' has usually been discussed
in association with phenomena closer to the sun, such as interchange reconnection \citep{FiskKasper20, schwadron2021switchbacks, bale2021solar} or some form of local dynamical activity \citep[e.g.,][]{ruffolo2020shear, SquireEA20}. Generally, polarity reversals at the heliospheric current sheet have been treated as a separate class. 

Recently, numerous studies have examined detailed properties of switchbacks in PSP data
\citep{ HorburyEA20,LakerEA21,MozerEA20,McManusEA20,Mozer21,TeneraniEA21}, while other studies have provided insight into both the possible origin of switchbacks \citep{LandiEA06} and their propagation and dissipation \citep{TeneraniEA20, magyar2021couldI, magyar2021couldII}.

The purpose here is to provide a more complete perspective of the occurrence rates of switchbacks by employing the data from the initial eight orbits of PSP in a unified analysis. For specificity, we will consider strong switchbacks that reverse the local polarity of the magnetic field, corresponding to deflection angles greater than $90^\circ$.

The paper is organized as follows: criteria for data selection and the definition of switchbacks are given in Sec.~\ref{sec:dataSB}; The statistics of SB waiting times and duration is presented in Sec.~\ref{sec:WTD}; The radial occurrence of SBs throughout the inner heliosphere is shown in Sec.~\ref{sec:radial}; The last section is dedicated to the discussion of the results. Finally, in the appendix, the SB database is presented.

%cite
%onlinecite
%textcite

\section{Data selection \& SB identification}
\label{sec:dataSB}

We use publicly available measurements of magnetic field from the MAG instrument on the FIELDS \citep{bale2016fields}, and solar wind ion speed from SPC on the SWEAP \citep{kasper2016solar} suites on the Parker Solar Probe \citep{fox2016solar} during its first 8 orbits. The analysis, from 2018 August 12 through 2021 June 19, is carried over non-overlapping 6-hour-long intervals as suggested in \citep{DudokDeWitEA20, bandyopadhyay2021energetic}. The following procedure has been automatized using the python library AI.CDAS that provides access to CDAS database through CDAS REST API. The API service provided by CDAS is an advantageous tool that allows a user to request data from the server and work with them dynamically. A 6h interval is considered for the analysis only if, when divided in six 1h-long subintervals, each subinterval has less than $\sim~30~\%$ missing points. In that case, linear interpolation is performed to restore data continuity (if any missing point is present). Intervals with more than $\sim~30~\%$ missing points have been discarded. The same criterion has been used to skim Wind measurements, over the same period, for the magnetic field \citep[MFI,][]{lepping1995wind} and bulk speed \citep[SWE,][]{ogilvie1995swe}). The detection of SBs was carried out using magnetic field measurements at 10s and 60s resolutions. To ensure that the choice of the above settings does not affect the results, the analyses have been repeated changing the interval length to 3 hours, and choosing a more strict limit of $15\%$ missing points per subinterval. We find a general consistency of results for the above variation of parameters. Therefore, the following analyses have utilized an interval duration of 6-hour and a $\sim 30 \%$ threshold in order to retain more intervals and a larger number of events.

The presence of switchbacks in each 6h interval, is revealed using the ``normalized deflection'' measure $z$ as defined by \citet{DudokDeWitEA20}
\begin{equation}
    z = \frac{1}{2} \left[ 1 - \cos(\alpha) \right],
    \label{eq:z}
\end{equation}
where $\alpha$ is the angle between the pointwise magnetic field $\BB$ and a local average $\langle \mathbf{\BB} \rangle$ evaluated over the considered 6 hours. ``Switchbacked'' regions are those with a value of $z \geq 0.5$ corresponding to magnetic field deflections larger than $90^\circ$ with respect to the local average. This is the main criterion employed in this study to define a switchback. This seems to us to be less arbitrary than a choice of other thresholds for $z$ that indicate non-reversing deflections.

\begin{figure}[ht]
    \centering
    \includegraphics[width=0.45\textwidth]{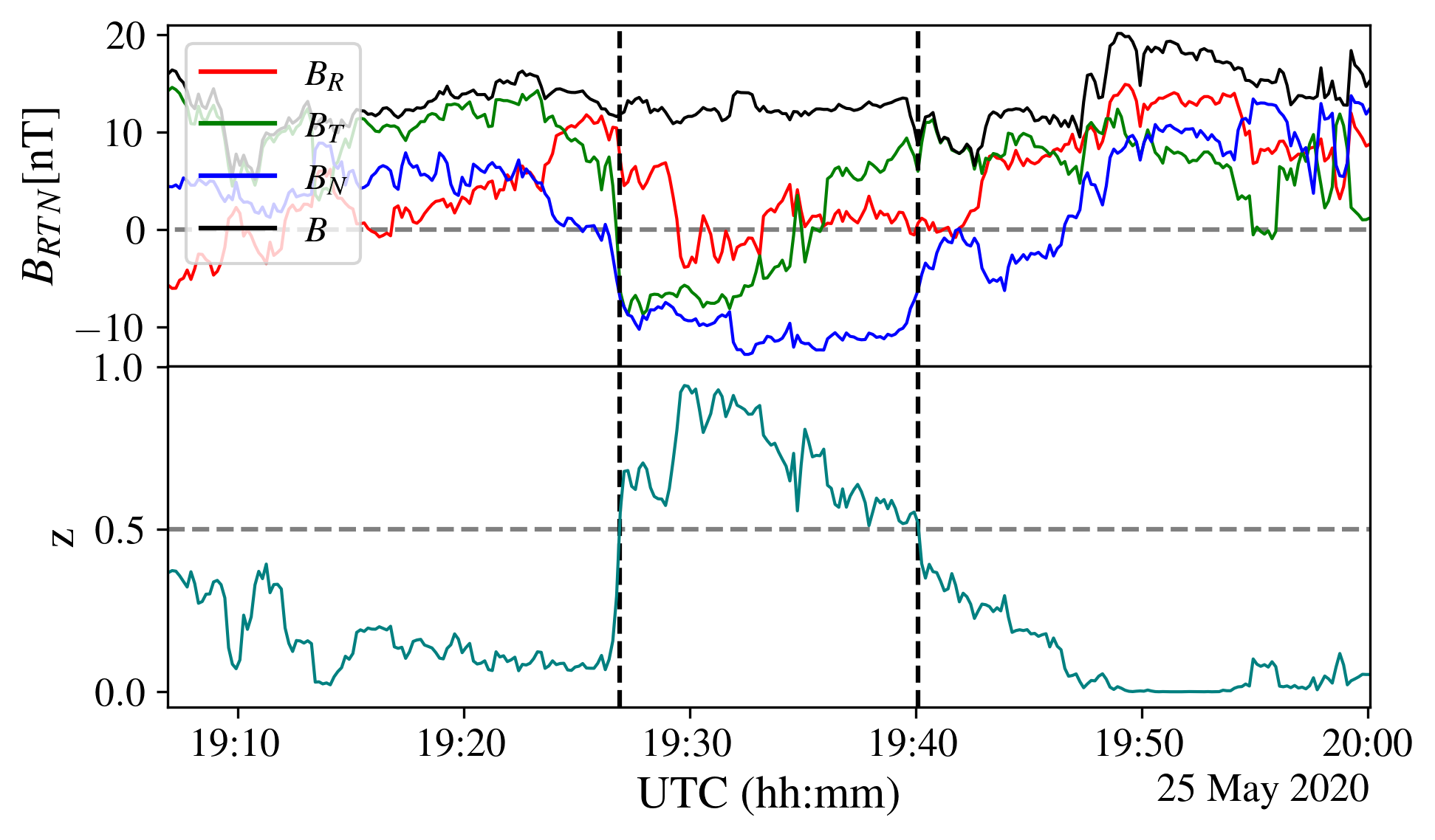}
    \caption{Example of a PSP time interval containing a switchback, as defined in Eq.~\ref{eq:z} and by \cite{DudokDeWitEA20}. (Top panel) magnetic field measurements at 10s resolution. (Bottom panel) The switchback, delimited by the two vertical dashed lines, is identified in the interval where $z \geq 0.5$, and extends for $\sim 13$ minutes.}
    \label{fig:SB_example}
    % WORKS/16_SB_database/orbits_1_8_10sSB_6hBavg/G_SB_profile_example_v02.py
\end{figure}

Figure~\ref{fig:SB_example}, shows an example of detected SB using PSP magnetic field data at 10s resolution. The $z$ parameter is constantly above the selected threshold of $0.5$ for $\sim 13$ minutes. Notice that the reversal, using the definition Eq.~\ref{eq:z} (as in  \cite{DudokDeWitEA20}), is identified using the total magnetic field vector, and not only one component.

\section{Waiting times and duration}
\label{sec:WTD}

The evaluation of the waiting time between two consecutive switchbacks and the duration of each event can provide insight to two fundamental questions: (I) do SBs tend to cluster, or are they more likely to be isolated events? (II) How do the waiting times and durations of switchbacks relate to turbulence scales, i.e., integral and inertial range scales? These questions address the issue of whether SBs are part of the evolving solar wind.

\begin{figure}[htp]
    \centering
    \includegraphics[width=0.35\textwidth]{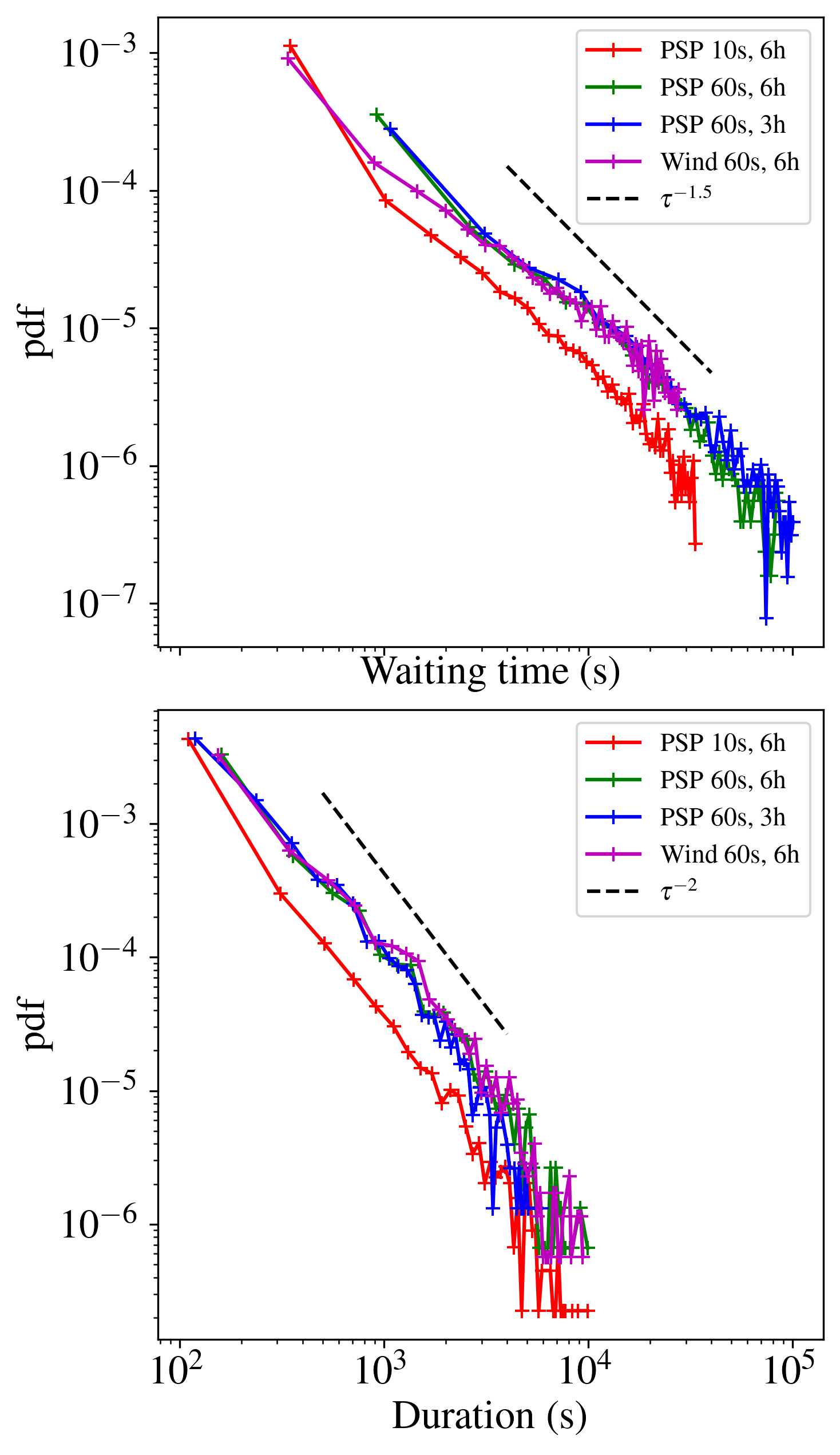}
    \caption{Distributions of (top) waiting times, and (bottom) SB duration, for the different tunable parameters -- magnetic field time resolution and averaging window extension -- used for PSP, and for Wind. Dashed lines are reference slopes.}
    \label{fig:WT_D_pdfs}
    % WORKS/16_SB_database/G_SB_WTs_duration_all_v00.py
\end{figure}

We define the duration of a SB as the number of consecutive above-threshold points times the cadence of magnetic field measurements employed (by this definition, a 1-point SB lasts 10s or 60s depending on the used dataset). In Fig.~\ref{fig:SB_example}, the duration of $\sim 13$ minutes is given by all the consecutive above-threshold events comprised between the two vertical dashed lines. Instead, the waiting time (WT) between SBs is defined as the time elapsed between the end of a SB and the beginning of the following. This last measure is necessarily affected by the presence of missing data intervals. To avoid spurious too-long waiting times, we discard waiting times with values greater than $\mu + 3\sigma$ --where $\mu$ is the mean of all the WTs and $\sigma$ is the standard deviation. Figure~\ref{fig:WT_D_pdfs} shows the distribution of WTs and duration for the different settings used to analyze PSP and Wind data, along with a slope of reference.

Meaningful differences and similarities are evident in the distributions reported in Fig.~\ref{fig:WT_D_pdfs}. With respect to the magnetic field resolution, all those obtained with 60s magnetic field overlap quite well. As will be discussed below, the only difference appears in the Wind waiting time distribution. The act of modifying the averaging window between 3 or 6 hours does not produce any appreciable variation. Instead, using different magnetic field cadences appear to be associated with appreciable changes in the distribution of waiting times. 

The distribution of SB duration shows a power law tail with a slope approaching $-2$. Therefore, an average duration can be estimated as $\langle \tau_{_{SB}} \rangle = \int_0^\infty t \; \mbox{pdf}(t) \; dt $, which varies between 200 and 500 seconds. Waiting time distributions, instead, show a slope closer to $-1.5$, for which the formal definition of an average WT is an ill-posed measure \citep{newman2005power}. However, since the distributions have a finite extension, an approximate (population) average WT, ranging from 18 to 90 minutes, can be calculated with the available measurements.

Table~\ref{tab:resume} shows a summary of the above-mentioned results and parameters used for the analysis. Different columns list the spacecraft name, the magnetic field cadence, the width of the window, the average switchback duration, and the average switchback waiting time.
The distribution of waiting times obtained with Wind measurements anticipates results to be discussed further in the next section. The maximum waiting time observed within all Wind measurements, is much lower than the average waiting time observed with PSP using the same cadence magnetic field (but also of the higher cadence). Therefore, we can infer that switchbacks are more common at 1~au since the average waiting time is shorter.

Consistent with previous discussions of switchbacks \citep{DudokDeWitEA20}
and discontinuities as measured by partial variance of increments (PVI)
\citep{chhiber2020clustering, BandyopadhyayEA20-PSPsep, sioulas2022statistical}, we note that a 
powerlaw waiting time distribution is suggestive of {\it clustering}. That is, these distributions imply the presence of correlation between successive switchbacks. 

\begin{table}[htp]
    \centering
    \begin{tabular}{ cccccc }
         SC & $\Delta t$ (s) & W (h) & $\langle \tau_{_{SB}} \rangle$ (s) & $\langle \tau_{_{WT}} \rangle$ (s) & $\langle \tau_{_{WT}} \rangle$ (min)  \\
         \hline
         PSP  &  60  &  6  &  450  & 4000 & 66 \\
         PSP  &  60  &  3  &  370  & 5500 & 91 \\
         PSP  &  10  &  6  &  200  & 1100 & 18 \\
         Wind &  60  &  6  &  503  & 2600 & 43 \\
         \hline
    \end{tabular}
    \caption{Switchback average duration ($\langle \tau_{_{SB}} \rangle$) and waiting times ($\langle \tau_{_{WT}} \rangle$), evaluated over the period 2018 August 12 through 2021 June 19, by varying the spacecraft (SC), the magnetic field resolution ($\Delta t$), and the averaging window width ($W$).}
    \label{tab:resume}
\end{table}

\section{Radial occurrence of switchbacks}
\label{sec:radial}

Early reports on switchbacks 
focused on PSP perihelia \citep{DudokDeWitEA20, MozerEA20}, 
and analyzed the first several orbits,
with somewhat contrasting selection criteria, and correspondingly 
diverse results
\citep{farrell2020magnetic, Mozer21, TeneraniEA21}. 
With the release of the data from PSP orbit 8, the pool of available measurements spans about three years ($1/3$ of the total nominal mission duration), and the study of the radial evolution of switchbacks becomes more accessible and statistically relevant.

More recent studies \citep{LiangEA21,MartinovicEA21,DudokDeWitEA20}
also report surveys, and 
vary rather significantly in both methodology
and criteria for selecting events.
To date, the origin of switchbacks remains 
a matter of debate.

\begin{figure}[htp]
    \centering
    \includegraphics[width=0.41\textwidth]{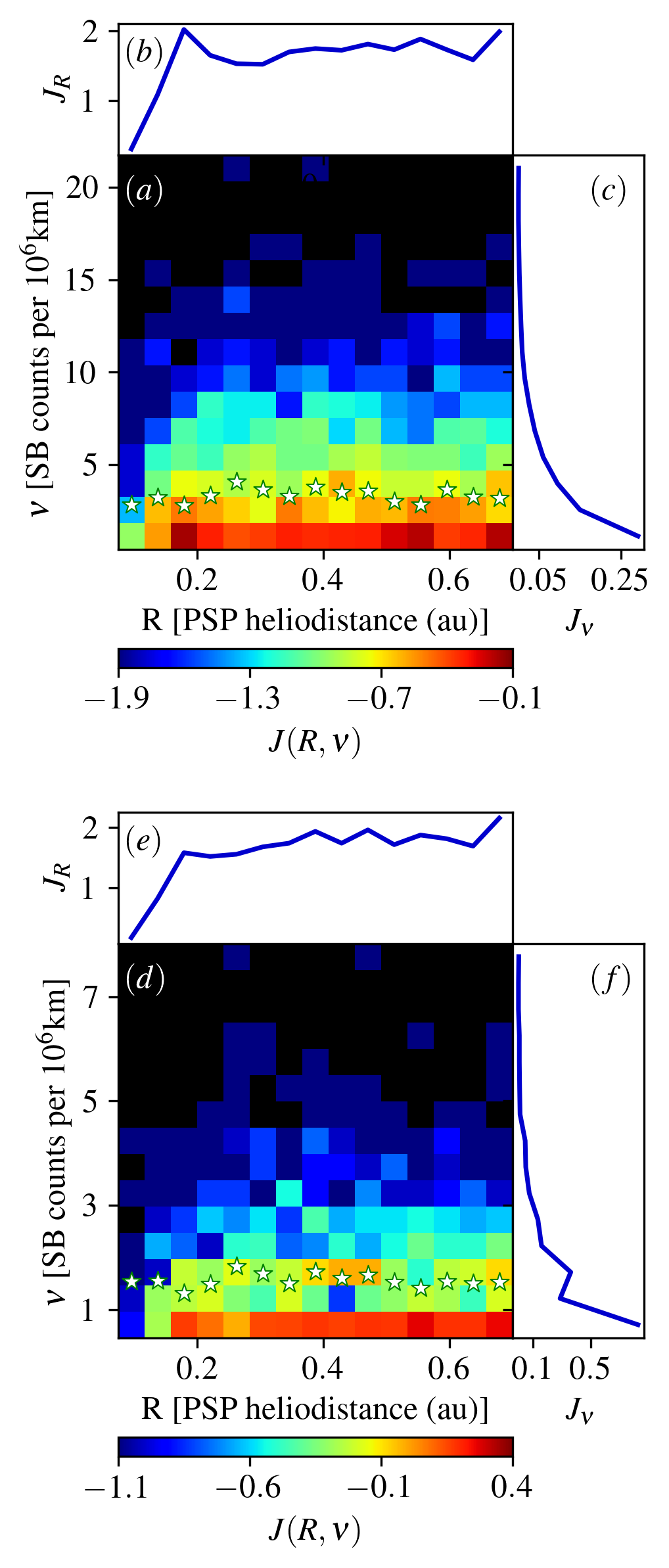}

    \caption{(Color plot) joint distribution $J(R,\nu)$ of SB counts per $10^6$~km $(\nu)$ vs PSP heliodistance $(R)$, using magnetic field at 10s (a) and 60s (d). Colors represent the probability distribution in log scale (black domains are regions with zero counts). The marginal probability distributions (see text) are shown in panels (b,c) for 10s and (e,f) for 60s magnetic field. White stars represent the average value $\langle \nu \rangle $ for each column, i.e. $\langle \nu \rangle = \frac{\int d\nu'~\nu'~F(\nu',R^*) }{ \int d\nu'~F(\nu',R^*) }$.}
    \label{fig:joint_pdf}
    % $\int d\nu'~\nu'~F(\nu',R^*) / \int d\nu'~F(\nu',R^*) $
\end{figure}

 Figure~\ref{fig:joint_pdf} shows a principal result of the present study -- the joint distributions of the occurrence of magnetic switchbacks per $10^6$~km vs PSP radial distance. The population in each bin is determined as follows: After having identified all the switchbacks as discussed in Sec.~\ref{sec:dataSB}, we count how many SBs occur in one hour, and record the radial distance of PSP at that time. As one would expect, the occurrence of SBs would depend on the speed of the solar wind. If the spacecraft sits in a stream of slow wind, the chance to measure SBs per unit time are lower than those referring to a fast stream. To overcome this trivial dependence, we make use of the Taylor hypothesis \citep{Jokipii73,chhiber2019contextual,PerezEA21} to convert time intervals to spatial distances employing SWEAP and SWE hourly averaged measurements (for PSP and Wind respectively). 
Therefore, we obtain an occurrence rate per unit distance for each 
one hour sample, and focus on 
this quantity instead of the rate per unit time.

The rates are tabulated from all available data, and are sorted in the binned parameter space $(R,\nu)$ -- where $R$ is PSP radial distance and $\nu$ is the number of SBs per Mkm.
When this has been accumulated over 
the entire 
population, we have obtained the joint distributions $J(R,\nu)$ that 
are shown in the color plots in panels (a, d) of Fig.~\ref{fig:joint_pdf}. The color parameter
is a measure of the logarithm of the probability density in a particular 
cell of the $(R,\nu)$ grid. 
An immediate impression is that 
a lower rate of SBs per km is 
is found at the smallest heliodistances.

A more direct view on the occurrence can be recovered via the marginal distributions that are obtained from the joint pdf. 
One marginal distribution is obtained by 
integration of the joint distribution over all radial distances, $J_\nu~=~\int J(R,\nu)~dR$. This is the distribution of SB counts per Mkm without regard to the radial distance. 

For the distribution computed from the 60s data,
noting the strong peak in $J_\nu$ in the range of $1< \nu <2$,
one draws a ``trivial'' interpretation that 
it is more common to find one or two SBs per Mkm
than it is to find several. A more physical interpretation of this result becomes apparent 
when recalling that the mean value of correlation lengths in the solar wind in the inner heliosphere gradually increases from about 500,000 km to 
1,000,000 km from 0.16 au to 1 au \citep{RuizEA14,ChhiberEA21-heliorad,cuesta2022isotropisation}.
Therefore, a simple interpretation of the 
substantial peaking of $J_\nu$
at values less than 2 for 60s magnetic field data,
is that, averaged over the aggregate population
within 1~au, switchbacks are found in
the solar wind about once per correlation length.
However, when 10s data is used in the calculation, 
the marginal distribution $J_\nu$ retains a similar 
shape, however stretched to a range of about $1 < \nu < 6$.
This is consistent with the intuition that higher time resolution data can (and will)
detect shorter time duration SBs that the coarser resolution data cannot detect. 

The other interesting
marginal distribution is $J_R~=~\int~J(R,\nu)~d\nu$, that is, the distribution of SBs per radial distance, as a function of the radial distance, and integrated over all possible local occurrence rates. This marginal distribution displays 
two distinct regimes: (I) descending below $0.2$~au $(\sim 40~R_\odot)$ there is a sharp decrease in high SB count rates, 
and (II) a plateau is reached between $0.2-0.7$~au (The available data from PSP at distances beyond 0.7 au are insufficient).

As we did in Sec.~\ref{sec:WTD}, we replicate and compare the same analysis with Wind measurements at 1~au. Since Wind's position is approximately fixed, it is not possible to reproduce the full joint pdf; but, below, we will show that the average occurrence rates at Wind are compatible with the PSP observations.

The average number of switchbacks per Mkm, as a function of the heliodistance, can be obtained from the joint distributions in Fig.~\ref{fig:joint_pdf}. Each column of the colour plot represents the distribution of the number of SBs per Mkm at a fixed heliodistance. From each of this conditioned distributions $F(\nu, R^*)$, it is possible to obtain the average $\nu$ for each fixed $R^*$, computing the normalized first moment $\int d\nu'~\nu'~F(\nu',R^*) / \int d\nu'~F(\nu',R^*)$. Each column average value (or first moment) is displayed with a star-shaped symbol in Fig.~\ref{fig:joint_pdf}.
The same treatment holds for Wind, except only one value of $R^*$ is present.

\begin{figure}[htp]
    \centering    
    \includegraphics[width=0.41\textwidth]{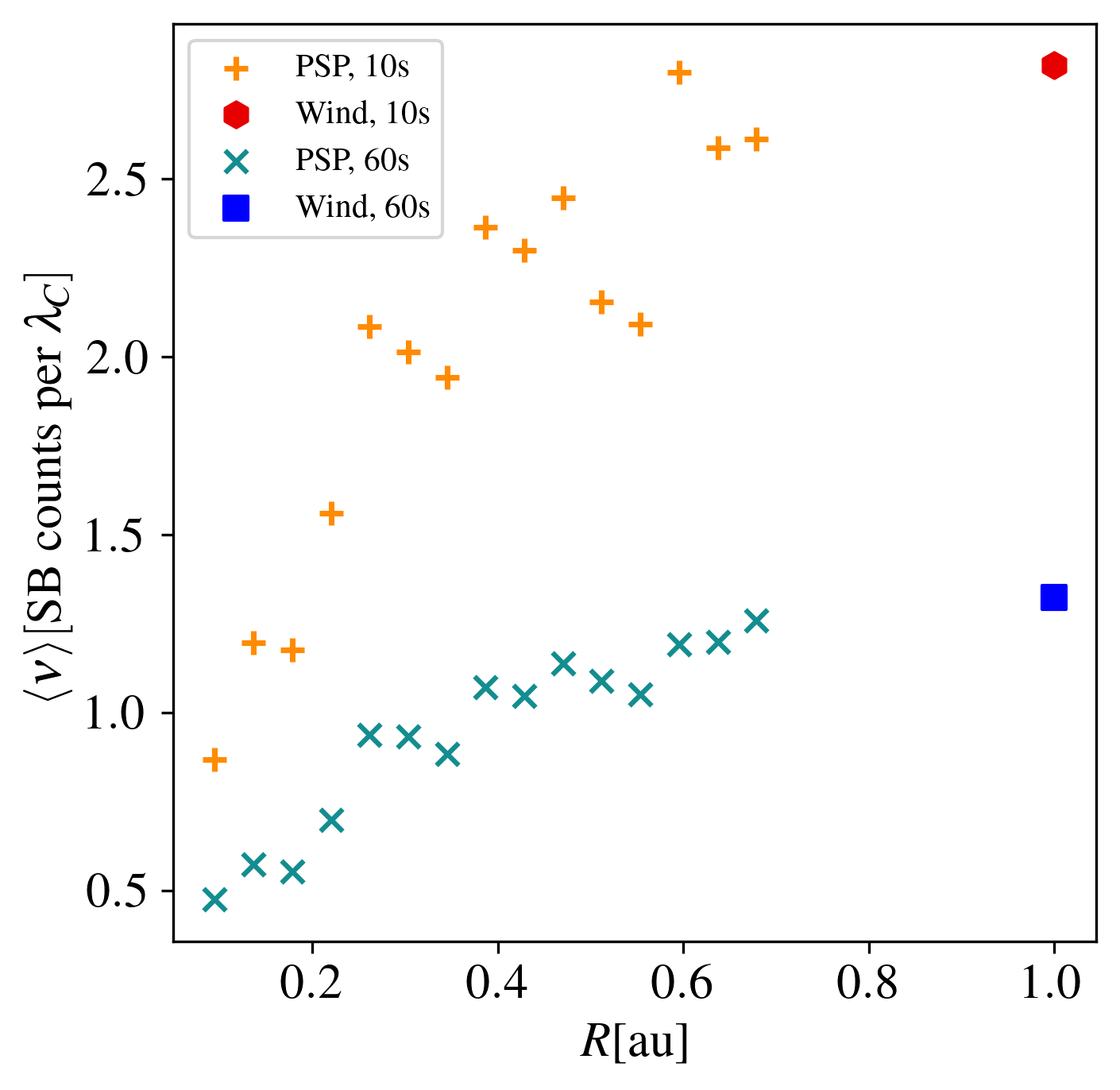}
    
    \caption{Average number of switchbacks per correlation length (estimated as $\lambda_C~\sim~10^6~\mbox{km} \sqrt{R(\mbox{au})}$), as a function of the heliodistance.}
    \label{fig:moneyplot}
\end{figure}

Again, we can notice that the first points (the four averages for $R \lesssim 0.25$~au) have values lower than the subsequent points, which seem to attain a plateau. An interesting physical interpretation arises if we consider the evolution of turbulence correlation length that can roughly be estimated as $\lambda_C~\sim~10^6~\mbox{km} \sqrt{R(\mbox{au})}$ \citep{RuizEA14,cuesta2022isotropisation}. After a simple transformation, we arrive at Fig.~\ref{fig:moneyplot} which shows the average number of switchbacks per correlation length as a function of the radial distance. As a check, the same analysis has been performed by measuring the average number of switchback directly from the detected population, rather than using the moments of the conditioned distributions, and the results (not shown) are indistinguishable.

The novel organization of the mean values, in Fig.~\ref{fig:moneyplot}, points toward the following consideration:
The average number of SBs per Mkm is constant, but the solar wind structure, in which they are embedded, is dynamically evolving.  If we consider the correlation length to be the typical scale of turbulent eddies, the result that the average number of SBs per correlation length increases, somehow envisions that SBs can be also seen as a byproduct of turbulence and their presence is enhanced as the turbulent cascade takes place.

%https://link.springer.com/article/10.1007/s11207-014-0531-9

\section{Discussion of the Results \& Conclusion}
\label{sec:discussion}

The topic of switchbacks in PSP observations has become very active 
due to the dramatic nature of these observations and the possible implications that these large magnetic-field deflections may have for unraveling the physics of solar wind heating and acceleration \citep{BaleEA19Nature,KasperEA19Nature},
which are central goals of the PSP mission.  Switchback observations, particularly the radial variations of their occurrence and other properties  studied by PSP, may be relevant 
to revealing the basic physics of 
the nascent solar wind. 

The prolific appearance of switchbacks in PSP datasets has prompted investigators to analyze {\it ensembles} of identified cases to establish statistical properties, and therefore, in principle, a robust physical perspective. These studies have adopted varying definitions, leading to disparate identification
criteria, thus emphasizing different physical properties. For example, the defining conditions in 
some studies do not require local polarity reversals, but only angular deflections \citep{DudokDeWitEA20}.
Other studies imply that high Alfv\'enicity should be a defining condition \citep{Borovsky16,HorburyEA18}.
Another approach is to identify source regions as a factor in the selection criteria \citep{TeneraniEA21}.
The approach employed here is intentionally simplified: we employ only the reversal of direction relative to a locally averaged field to be the sole fundamental condition for identifying a  switchback. 
As an a posteriori check, we examined the fraction of our identified events to assess how many are likely to be 
associated with 
possible heliospheric current sheet crossings; the 
results indicate such 
current sheet crossings represent 
$\sim 2\%$ of the identified PSP events, and $\sim 0.3\%$ for the Wind events. Therefore, these cases do not have a substantial effect on the statistical descriptions
we have provided. 

There are of course subsidiary conditions that enter into the selection we have implemented. For example, as discussed in Section \ref{sec:dataSB} the average magnetic field must be well-defined within the interval in question, and the appearance of too large a fraction of missing data points renders a particular time slice unusable. Our methodology, applied between 2018 August 12 and 2021 June 19, about 1042 days, yielded a total of 22,223 switchbacks using PSP 10 second data and 6 hour averaging interval. See Table~\ref{tab:sbs_number} for a list including all configurations used in this study.

\begin{table}[htp]
    \centering
    \begin{tabular}{ c c c c  }
        $\Delta t$(s) & W(h) & \#SBs (PSP) & \#SBs (Wind)  \\
         \hline
         10 & 3  & 19141   & 23793  \\
         10 & 6  & 22223   & 29908  \\
         60 & 3  & 6458    & 8117   \\
         60 & 6  & 7556    & 9301   \\
         \hline
    \end{tabular}
    \caption{Total number of switchbacks \#SBs identified with PSP and Wind, 
    varying the magnetic field cadence $\Delta t$ in seconds 
    and the averaging window width $W$ in hours. }
    \label{tab:sbs_number}
\end{table}

We note that we included analyses only of data at heliocentric distances $r < 0.7$~au,
even though some data was available in the range $ 0.7~\mbox{au} < r< 1~\mbox{au}$.
That range is excluded due to the relative scarcity of data. In part this is 
due to the PSP orbit aphelia, which barely extend 
beyond 0.7~au after orbit 6, as well as lower data availability in general in that range
\citep{ChhiberEA21-heliorad}.

{\it Characteristics of the observed switchback rates.}
A salient feature of the switchback occurrence rates 
seen in both Fig.~\ref{fig:joint_pdf}
and Fig.~\ref{fig:moneyplot}
is the appearance of a ramp at $r< 0.25$~au (i.e., $\sim$48~\Rsun).
A naive extrapolation of this trend 
to zero switchback rate occurs at around 20 solar radii.
This is remarkably close to the radius at which the 
first passage into subAlfvénic coronal plasma was detected 
recently \citep{KasperEA21-prl, chhiber2022extended}.
If one views this as an increase in switchback rate 
beginning at $\sim 20$ solar radii, then the 
``turning on'' of switchbacks in this range of heliocentric distances
appears to support the hypothesis that shear driven activity is initiated in this vicinity.
A proposed candidate is MHD mixing layer dynamics, leading to rollups of vorticity and magnetic field reversals \citep{ruffolo2020shear}. This is also the same region in which 
the more striated coronal plasma transitions to 
a more isotropic {\it flocculated} appearance in heliospheric imaging 
analyses \citep{DeForestEA16,ChhiberEA19-1}
of the inner solar wind.

Beside the \citet{ruffolo2020shear} model, 
there are other models for {\it in situ} generation of switchbacks 
\citep{SquireEA20,schwadron2021switchbacks} that may also be consistent 
with the present observations. We also are not attempting here to 
garner evidence to 
controvert models for generation of switchbacks
at much lower altitudes, such as 
interchange reconnection in the chromosphere or lower corona
\citep{FiskKasper20,drake2021switchbacks,magyar2021couldI,magyar2021couldII,zank2020origin}. In fact, 
there may be separate 
populations generated at lower altitudes 
that experience attenuation \citep{TeneraniEA20}
during propagation to radii beyond
the Alfvén transition zone, and a separate population
that is initiated at higher altitudes \citep{TeneraniEA21}.

It is noteworthy that transforming the data from a switchbacks per kilometer format to a switchbacks per correlation scale format, as in Figure~\ref{fig:moneyplot}, does appear to produce a systematic ordering of the data. For example, the trends at low heliocentric distance of 
switchbacks/$\lambda_c$  for both magnetic cadences 
approach a zero count rate near 0.1 au.

Meanwhile, the relatively smooth increase at larger $R$ extrapolates reasonably well to the Wind results at 1~au. It is tempting to interpret this as support for the idea that turbulence properties are, at some level, controlling or influencing switchback rates.
However, it is difficult to assert that scaling to the correlation length is the only way to organize the data.

Albeit the driving of the source in the lower solar atmosphere is definitely responsible for defining some of the macroscopic features of turbulence \citep{bale2021solar, fargette2021characteristic, magyar2021couldI, magyar2021couldII}, given the systematic behavior of the switchback rate seen in Figure~\ref{fig:moneyplot}, there is a strong suggestion that switchback properties, including their rate of occurrence, are evolving and are controlled, at some level, by dynamics outside of about 0.1 au.

\begin{acknowledgments}

This research is partially supported by the Parker Solar Probe mission through the IS$\odot$IS Theory and Modeling team and a subcontract from Princeton University (SUB0000317),  NASA PSP Guest Investigator grants 80NSSC21K1765 and 80NSSC21K1767, and NASA Heliospheric Supporting Research program grants
80NSSC18K1210 and 80NSSC18K1648. This project is also partially supported by the European Union's Horizon 2020 research and innovation program under grant agreement No. 776262 (AIDA, www.aida-space.eu). The PSP data used here is publicly available on NASA CDAWeb \url{https://cdaweb.gsfc.nasa.gov/index.html/}.
The authors thank the FIELDS team (PI: Stuart D. Bale, UC Berkeley), the Integrated Science Investigation of the Sun (\isois) Science Team (PI: David McComas, Princeton University) and the Solar Wind Electrons, Alphas, and Protons (SWEAP) team for providing data (PI: Justin Kasper, BWX Technologies) for providing data. Parker Solar Probe was designed, built, and is now operated by the Johns Hopkins Applied Physics Laboratory as part of NASA’s Living with a Star (LWS) program (contract NNN06AA01C). Support from the LWS management and technical team has played a critical role in the success of the Parker Solar Probe mission.
FP thanks Mattia Rovito for his help with the design of the database webpage.
LP thanks the ICT center of the University of Calabria for hosting the webserver of the database.

\end{acknowledgments}

\appendix

\section{Online database of detected switchbacks}
\label{appendix}

The present work afforded an opportunity to assemble an open-access online catalogue that accepts inquiries in the form of a specified time interval, and provides a list of magnetic field reversals and the time of their occurrence. Assembly of such a database requires analyzing a large amount of raw data (months or years), from the public PSP archive folders.

The reproduction of such data can be achieved by replicating the download and analysis steps; this sacrifices quickness in favor of economy of storage requirements. Another approach is to set up a capacious disk where storing all possibly needed data sets to access them rapidly using a simple interface. The assembled database supports the latter approach for convenience of the interested user.

The database is built using the parameters described in the paper. The user is asked to input the spacecraft (Wind or PSP), magnetic field cadence (10s or 60s), averaging window (3h or 6h), and a time interval (within the first 8 orbits of PSP). The output is immediately provided as a list of UTC seconds, human-readable dates, and the value of $z$, for all the points that satisfy $z \geq 0.5$. The user can then choose to print the output on screen, or download as a text file.

The database is freely accessible at \url{http://astroplasmas.unical.it/SBDB/}

%\end{multicols}

%\nocite{*}

\bibliography{biblio,
ag,hl,mp,qz,refs_WHM}% Produces the bibliography via BibTeX.

\end{document}